\documentclass{article}
\usepackage[T1]{fontenc}
\usepackage[latin9]{inputenc}
\usepackage[letterpaper]{geometry}
\usepackage{textcomp}
\usepackage{amstext}
\usepackage{graphicx}
\usepackage{esint}
\usepackage[authoryear]{natbib}

\makeatletter
\newcommand{\lyxaddress}[1]{
	\par {\raggedright #1
	\vspace{1.4em}
	\noindent\par}
}

\makeatother

\begin{document}
\title{Analysis of Jupiter's deep jets combining Juno gravity and time varying magnetic field measurements}
\author{Keren Duer$^{1}$, Eli Galanti$^{1}$
 and Yohai Kaspi$^{1}$ \\
\\
(Astrophysical Journal Letters, in press)}
\maketitle

\lyxaddress{\begin{center}
\textit{$^{1}$Department of Earth and Planetary Sciences, Weizmann
Institute of Science, Rehovot, Israel. }\\
\par\end{center}}

\begin{abstract}
\textbf{Jupiter's internal flow structure is still not fully known, but can
be now better constrained due to Juno's high-precision measurements.
The recently published gravity and magnetic field measurements have
led to new information regarding the planet and its internal flows,
and future magnetic measurements will allow taking another step in
resolving this puzzle. In this study, we propose a new method to better constrain Jupiter's internal flow field using the Juno gravity measurements combined with the expected measurements of magnetic secular variation. Based on a combination of hydrodynamical and magnetic field considerations we show that an optimized vertical profile of the zonal flows that fits both measurements can be obtained. Incorporating the magnetic field effects on the flow better constraints the flow decay profile. This will allow getting closer to answering the long-lived question regarding the depth and nature of the flows on Jupiter.}
\end{abstract}

\section{Introduction\label{sec:Introduction} }

The nature of Jupiter's interior is still a great mystery. The recent
discovery that the depth of Jupiter's surface winds is $\sim3000$
km \citep{Kaspi2018} raises the possibility that the flow penetrates
to depths where the electrical conductivity is large enough so that
the flow might interact with the magnetic field. Due to Jupiter's
large mass and density, the inner pressure is high enough to cause
gas ionization relatively close to its surface. The significant ionization
is expected at $\sim0.97$ of the radius of Jupiter ($\sim2000$ km
from its surface), getting stronger with depth \citep{Liu2008,French2012}.
According to the Juno gravity field measurements \citep{Kaspi2018}
the flow is expected to penetrate the ionized region and therefore
influence the magnetic field and vice versa. 

The electromagnetic (EM) induction equation describes the temporal
variability of the magnetic field due to interaction with the flow,
the strength of the magnetic field itself and the electrical conductivity
of the fluid. Since the electrical conductivity at Jupiter's higher
atmosphere (cloud-level to $\sim2000$ km) is very low \citep{Liu2008,French2012},
the jet streams are expected to have no interaction with the magnetic
field. In this region, the flows have a significant effect on the
gravity field \citep{Kaspi2013a,Kaspi2018}. The inner region, below
$\sim2000$ km with remnants of jet velocities, is associated with
an electrical conductivity that increases exponentially with depth,
and might affect magnetic field anomalies \citep{Galanti2017c}. Note
that the transition between the two regions is likely gradual and
not abrupt. 

Several studies examined the possible interaction between the flow
and the magnetic field in gas giants \citep[e.g.,][]{Liu2008,Gastine2012,Jones2014,Cao2017,Dietrich2018},
but none have used the gravity and magnetic field measurements simultaneously
to add constraints to the possible flow profile in Jupiter and other
gaseous planets. Previous studies that estimated the flow decay structure
have not considered the interaction between the flow and magnetic
fields \citep[e.g.,][]{Kaspi2018}. One study presented the mean-field
electrodynamic balance (MFED) as a method for constraining the deeper
regions of the decay flow profile \citep{Galanti2017c}, but it is
likely that due to the Juno measurements showing strong azimuthal variations
in the magnetic filed, the magnetic field of Jupiter is too complicated
for using the MFED method. Use of this method will result in different
vertical structure for each longitude, which is highly unconstrained.
Conversely, the time changes in the magnetic field (also named magnetic
secular variation), are likely to be measurable within the upcoming
Juno measurements \citep{moore2018}, and might be used to better
constrain the flow field.

Temporal variation in the magnetic field generated by the core dynamo
has been also identified for the Earth \citep[e.g.,][]{vestine1966,kahle1967,Bloxham1985},
and estimated for other planetary interiors such as Jupiter's \citep[e.g.,][]{Gastine2014,Jones2014,ridley2016}.
On Earth, the secular variation (SV) is exploited to yield flow profiles
at the top of Earth's outer core \citep{Bloxham1985,bloxham1992}.
While on Earth the SV strongly indicates internal flows in the
core \citep{holme2007}, in Jupiter the SV can be a result of multiple
origins, among them interior flows in the fully conductive region, external sources and zonal flows in the semi conductive region
\citep{Gastine2014,ridley2016}. SV time scale associated with zonal
flows for Jupiter's magnetic field is estimated to be notable even
within a single year \citep{Gastine2014,moore2018,moore2019}, if
measured with high enough resolution. \citet{moore2019} show that
time variations in Jupiter's internal magnetic field recorded from
different spacecrafts (from Pioneer 10 until Juno) over the past $50$
years are consistent with advection by zonal flows. However, since
Juno revealed a complex longitudinal structured magnetic field, it
is possible that using all passes of the Juno mission itself will allow tracking
the advection of these longitudinal features and better estimation
of the vertical profile of the zonal flows.

In this study, we use the recently published gravity measurements
from Juno \citep{Iess2018}, together with the magnetic field measurements
\citep{Connerney2018} and the upcoming time-varying magnetic field
measurements to better understand the nature of the deep flows in
Jupiter. We present a method for the calculation of the decay profile
of Jupiter's surface winds that can explain both the Juno gravity
measurements and the time-varying magnetic measurements, which are currently measured by Juno. The latter are simulated
and used to constrain the lower region of the flow decay profile by
the EM induction equation that relates the magnetic secular variation
to the flow strength. The upper region associated with strong flows
is determined by relating the gravity field to the flows via thermal
wind balance \citep{Kaspi2010a}. We also characterize the transition
between the regions using a variety of flow decay options and multiple
transition options to find the best solution that can fit both measurements. 

The manuscript is organized as follows: in section \ref{sec:Methodology}
we present the relation connecting the flow field and the gravity
(aka thermal wind (TW) balance), and the magnetic secular variation (MSV)
model, and describe the simulation and optimization processes. In section
\ref{sec:Results} the combined solution based on both gravity and
magnetic fields is presented, and in section \ref{sec:Conclusions}
we discuss the implications of this study and conclude.

\section{Methodology\label{sec:Methodology}}

In order to better constrain the deep flow on Jupiter, we combine
two fundamentally different approaches. The first is based purely
on the gravity measurements and their relation to the flow via TW
balance \citep[e.g.,][]{Kaspi2013a,Galanti2017c,Kaspi2018}, and the
other is based on the EM induction equation called the MSV method
\citep{Bloxham1985,bloxham1992}. We aim to use the gravity constraints
to find the best solution for the upper region of the flow, and the
MSV method to better constrain the lower region of the flow and the
transition between the regions.

\subsection{Gravity field constraints\label{subsec:Gravity-field}}

Jupiter's large size and rapid rotation imply that the leading order
momentum balance is geostrophic, namely a balance between the Coriolis
force and the horizontal pressure gradients \citep{Vallis2006}. Consequently,
the leading order vorticity balance is thermal wind balance \citep{Kaspi2009}.
The zonal (azimuthal) component of the thermal wind balance is 
\begin{equation}
2\Omega\frac{\partial}{\partial z}\left(\tilde{\rho}u\right)=\tilde{g}\frac{\partial\rho^{\prime}}{\partial\theta},\label{eq:thermal wind}
\end{equation}
where $\Omega$ is the planetary rotation rate of Jupiter ($\frac{2\pi}{\Omega}\cong9.92$
h), $z$ is the direction parallel to the spin axis, $u\left(r,\theta\right)$
is the zonal flow, with $\theta$ being latitude and $r$ is the depth,
$\tilde{\rho}\left(r\right)$ is the background radially dependent density field,
$\tilde{g}(r)$ is the corresponding radial gravitational acceleration
and $\rho'\left(r,\theta\right)$ is the density anomaly related to
the zonal flow. Expanding to higher order balances, beyond thermal
wind, including the contributions due to oblateness, is possible,
but it has been shown that for determination of the deep flows, thermal
wind balance (Eq.~\ref{eq:thermal wind}) is the leading order balance
and is sufficient \citep{Galanti2017a}.

The zonal gravitational harmonics, induced by the dynamics, are calculated
by integrating the density anomaly $\rho^{\prime}$ from Eq.~\ref{eq:thermal wind}
\citep{Kaspi2010a}. They are represented by 
\begin{equation}
\Delta J_{n}^{{\rm mod}}=-\frac{2\pi}{MR_{J}^{n}}\intop_{0}^{R_{J}}r^{n+2}dr\intop_{s=-1}^{1}P_{n}\left(s\right)\rho^{\prime}\left(r,s\right)ds,\label{eq:Jn}
\end{equation}
where $\Delta J_{n}^{{\rm mod}},\;n=2,...,N$ are the harmonic gravity
coefficients induced by the zonal flows, $M$ and $R_{J}$ are Jupiter's
mass and radius, respectively, $P_{n}$ are the associated Legendre polynomials
and $s=r\cos\left(\theta\right)$. The gravity harmonics can be used
to calculate the latitude-dependent gravity anomalies in the radial
direction $\Delta g_{r}^{{\rm mod}}\left(\theta\right)$, so representation
of the model results is available in both terms \citep{Kaspi2010a,galanti2017e}.

Giant planets are, to leading order, north-south symmetric bodies.
The value of the low-degree even gravity harmonics reflect mostly
the internal mass distribution within the planet, resulting from the
planet's shape and rotation \citep{Hubbard2012}. Therefore, the even
degrees ($n=2,4,6...$) resulting from the flow are difficult to be
differentiated from the total values and might not be a good indicator
for the dynamics \citep{Kaspi2013a,Kaspi2017}. If no north-south
asymmetry exists, the odd $J_{n}$ should be identically zero. Juno
was able to measure all the gravity harmonics within the sensibility
range \citep{Iess2018,Kaspi2018}, and found significant non-zero
values for the odd gravity harmonics. Since rigid body rotation cannot explain
these values, wind asymmetry between the hemispheres must be the cause
for the odd $J_{n}$. Consequently, for the odd harmonics $\Delta J_{n}=J_{n}$.
Following \citet{Kaspi2018}, we use the TW approach to calculate
the flow decay profile that can explain the four odd gravitational
moments ($J_{3,\:}J_{5,\:}J_{7,}$ and $J_{9}$) within the uncertainty
range.  Unlike \citet{Kaspi2018}, we also require that this profile
is physically consistent with magnetic field considerations as described
below.

\subsection{Time dependent magnetic field constraints\label{subsec:Time-dependent-magnetic}}

The temporal variation of Jupiter\textquoteright s magnetic field
can be described by the Maxwell equations, combined to set the EM
induction equation \citep[e.g.,][]{Jones2011},\textbf{ 
\begin{equation}
\frac{\partial\mathbf{B}}{\partial t}=\nabla\times\left(\mathbf{u}\times\mathbf{B}\right)-\nabla\times\left(\eta\nabla\times\mathbf{B}\right),\label{eq: induction}
\end{equation}
}where $\mathbf{B}$ is the three dimensional magnetic field, $\mathbf{u}$
is the three dimensional velocity and $\eta=\frac{1}{\mu_{0}\sigma}$
is the magnetic diffusivity, with $\mu_{0}$ being the magnetic permeability
of free space and $\sigma\left(r\right)$ is the electrical conductivity. 

If Jupiter's internal flow indeed extends down to regions of high
electrical conductivity, a non-negligible effect on the time dependent
magnetic field is expected. This effect is not the only one that can
cause magnetic secular variation, and other effects such as an external
magnetodisk \citep{ridley2016}, internal dynamo variabilities \citep{Jones2014,Dietrich2018} or
other rotation periods \citep{moore2019} are possible. However,
within the time scales we are addressing, zonal flow advection is
the predominant one. Considering only steady zonal flows are considered, the
non-linear interaction term between the flow field and the magnetic
field in the radial direction becomes 
\begin{equation}
-\frac{u}{r\sin\theta}\frac{\partial B_{r}}{\partial\phi},\label{eq: radial magnetic}
\end{equation}
where $\phi$ is longitude, $\theta$ here is co-latitude and $\frac{u}{r{\rm sin}\left(\theta\right)}$
is the angular velocity associated with the zonal flows. The magnetic
field can be decomposed into poloidal and toroidal components such
that 
\begin{equation}
\mathbf{B}=\nabla\times\left(\nabla\times P\mathbf{e}_{\mathbf{r}}\right)+\nabla\times T\mathbf{e}_{\mathbf{r}},\label{eq: Br decomposition}
\end{equation}
where $P$ and $T$ are the poloidal and toroidal potentials, respectively.. 

Similar to the gravity decomposition, the EM induction equation is
solved with a pseudo-spectral method. Using spherical harmonics in
the horizontal directions and Chebyshev polynomials in the radial
direction, the poloidal potential is then 
\begin{equation}
P=\sum_{n}\sum_{l}\sum_{m}P_{lmn}C_{n}\left(r\right)Y_{l}^{m}\left(\theta,\phi\right),\label{eq:grlm}
\end{equation}
where $C_{n}\left(r\right)$ are the Chebyshev polynomials of degree
$n$, $Y_{l}^{m}\left(\theta,\phi\right)$ are the spherical harmonics
of degree $l$ and order $m$, and $P_{lmn}$ are the coefficients
associated with each combination of the Chebyshev and spherical harmonic
functions. For the toroidal potential an equivalent equation is used.
Note that representation of the potentials is possible also with the Schmidt
coefficients as in \citet{Connerney2018}. The poloidal magnetic potential
governing equation then becomes 
\begin{equation}
\frac{l\left(l+1\right)}{r^{2}}\left[\left(\frac{\partial}{\partial t}+\lambda\frac{l\left(l+1\right)}{r^{2}}\right)C_{n}-\lambda C_{n}^{\:\prime\prime}\right]P_{lmn}=-\intop Y_{l}^{m*}\frac{u}{r\sin\theta}\frac{\partial B_{r}}{\partial\phi}d\mathbf{\Omega}.\label{eq: poloidal governing}
\end{equation}
where the double prime denotes second derivative in $r$, $\lambda=\frac{\eta}{\eta_{{\rm bottom}}}$
is the normalized magnetic diffusivity and $\mathbf{\Omega}$ is the
solid angle. Juno was able to measure
the three components of the magnetic field outside of Jupiter (the
scalar potential field) in the first year of orbit \citep{Connerney2018},
which can be used to construct a map of the averaged $B_{r}$ during
the first year of measurements (the Juno reference model through Perijove
9 - JRM9). In the following orbits, Juno will continue to measure
the scalar potential of Jupiter's magnetic field to yield a new map
of the averaged $B_{r}$ during this time. If there will be a notable
change in the spatial structure of $B_{r}$ between the two periods,
the induction relation can be used to
determine the vertical profile of the flow. We search for a flow decay
profile that will generate changes in the magnetic field (and therefore
in $P$ and $T$) that will best fit the measured time changes in
\textbf{$B_{r}$}.

\subsection{Simulation and measurements\label{subsec:Simulation-and-measurements}}

Since the gravity measurements are already available \citep{Iess2018},
while the temporal variation in the magnetic field is not, validation
of the solution is possible only with respect to one of the two measurements
that we aim to explain. In order to test the methodology presented
here, we simulate the time varying magnetic field measurements. The
simulated measurements are generated by integrating the induction
equation (Eq.~\ref{eq: induction}) with a chosen flow structure
($\mathbf{u}$ in Eq.~\ref{eq: induction}) for approximately one
year.

Using the decay profile of \citet{Kaspi2018} is found to be inconsistent
with the time changing magnetic field. The resulting time variation
in the magnetic field, after integrating the induction model for one
year, is too large and not physical
(not shown). This by itself indicates that the \citet{Kaspi2018} profile
is too strong in regions of high electrical conductivity and hence
cannot be used to constrain the flow at depths below $\sim3000$ km.
Moreover, any solution that includes substantial deeper zonal flows
will be inconsistent with the magnetic field constraints \citep[e.g.,][]{Kong2018}.

\begin{figure}[h]
\begin{centering}
\includegraphics[width=0.7\textwidth]{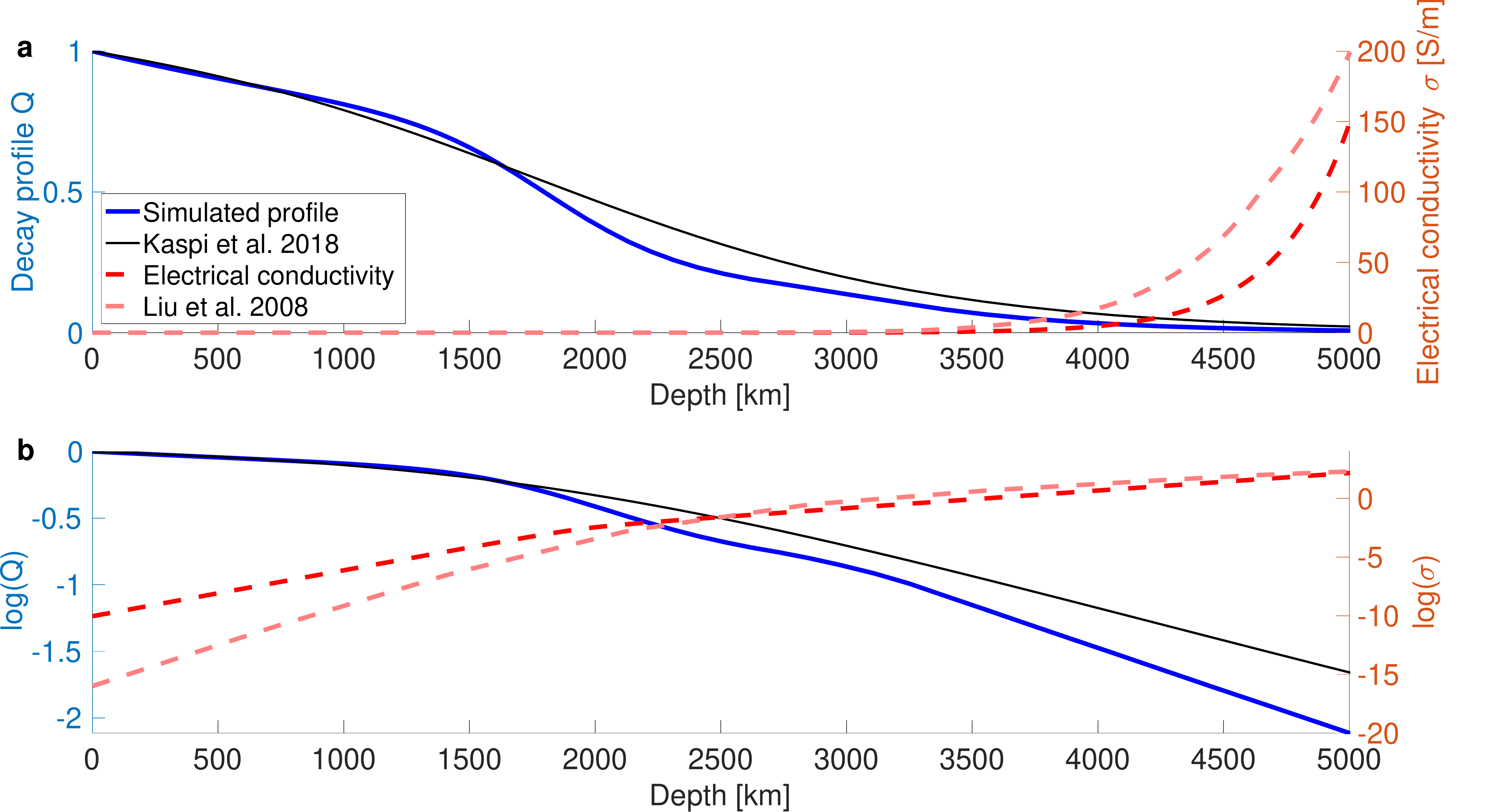}
\par\end{centering}
\caption{\label{fig:decay and conductivity}The decay profile of \citet{Kaspi2018}
(black), the simulated decay profile used to generate the $\frac{\partial B}{\partial t}$
measurements (blue), the electrical conductivity (dashed red) and
the electrical conductivity as appears in \citealt{Liu2008} (light
dashed red) in linear (a) and log scales (b).}
\end{figure}

Therefore, we define a modified decay profile that is similar to the
\citet{Kaspi2018} profile above $\sim2000$ km and weaker below (Fig. \ref{fig:decay and conductivity}),
yet still generates reasonable gravity harmonics \textbf{ }(Table
1). The induction model is then integrated with the new profile to
create ``measurements'' of the time varying magnetic field according
to Eq.~\ref{eq: induction}-\ref{eq: poloidal governing}. We set
this decay profile with a relatively small number of parameters, but
with enough freedom to contain the needed complexity. The upper region
is similar to \citet{Kaspi2018}, and the lower region is a simple
exponential decay, so that the overall simulated flow structure is

\begin{equation}
U_{s}(\theta,r)=U_{{\rm surf}}(s)Q_{s}(r),\label{eq:simulation wind}
\end{equation}
\begin{equation}
Q_{s}(r)=(1-\alpha)\exp\left(\frac{r-R_{J}}{H_{1}}\right)+\alpha\left[\frac{{\rm tanh}\left(-\frac{R_{J}-H_{2}-r}{\Delta H}\right)+1}{{\rm tanh}\left(\frac{H_{2}}{\Delta H}\right)+1}\right]\qquad R_{T}\leq r\leq R_{J},\label{eq:simulation decay upper}
\end{equation}
\begin{equation}
Q_{s}(r)=Q_{s}(r=R_{T})\exp\left(\frac{r-R_{T}}{H_{3}}\right)\qquad r<R_{T},\label{eq:simulation decay lower}
\end{equation}
where $U_{{\rm surf}}(s)$ is the measured wind at $R_{J}$ \citep{Tollefson2017}
projected toward the planet's interior in the direction parallel to
the spin axis, $Q_{s}(r)$ is
the radial decay simulated function (Fig. \ref{fig:decay and conductivity}),
and the set of parameters that forms the decay rate are: $\alpha=0.5$,
$H_{1}=2389$, $H_{2}=1830$ km, $\Delta H=500$ km, $R_{T}=0.958R_{J}$
and $H_{3}=660$ km.

Interestingly, it appears that $J_{3}$ is very sensitive to depths
of $2500-4000$ km and that a weak flow at those depths does not allow
fitting $J_{3}$ to the Juno measurements. Combined with the fact
that the induction model requires very weak flow below $3500$ km,
in order to generate reasonable changes in the magnetic field,
we find that $J_{3}$ constrains the flow to reach deeper than $3500$
km, but the flow at those depths cannot be higher than $\sim5-10\,{\rm ms^{-1}}$,
and can reach no more than $\sim1\,{\rm ms^{-1}}$ at $5000$ km.

Due to Jupiter's large mass, its electrical conductivity is relatively high
close to its surface. To date, Jupiter's electrical conductivity profile
in the upper $5000$ km of the planet is based mostly on ab initio
simulations and shock wave experiments \citep*{nellis1992,nellis1995,Weir1996,Liu2008,French2012}.
The latest published magnetic field \citep{Connerney2018} combined
with the induction model in this study, requires relatively low conductivity
at depths of $2000-5000$ km, to allow reasonable changes in the magnetic
field with time. The conductivity profile chosen here is similar to the profile
of \citet{Cao2017}, and is within the error range of the \citet{Liu2008}
profile (Fig. \ref{fig:decay and conductivity}). We model the magnetic
diffusivity (the inverse of the electrical conductivity) similar to
\citet{Jones2014,Cao2017} and \citet{Dietrich2018} such that: 
\begin{equation}
\eta(r)={\rm F\cdot exp}\left(c+\sqrt{c^{2}+d}\right)
\end{equation}
where $c=\frac{1}{2}[(g_{1}+g_{3})r-g_{2}-g_{4}],\: d=(g_{1}r-g2)(g_{3}r-g_{4})-g_{5},\:F=10,\:g_{1}=645.81,\:g_{2}=611.42,\:g_{3}=246.03,\:g_{4}=222.54$ and $g_{5}=0.21919$
.

\begin{figure}[h]
\begin{centering}
\includegraphics[width=1\textwidth]{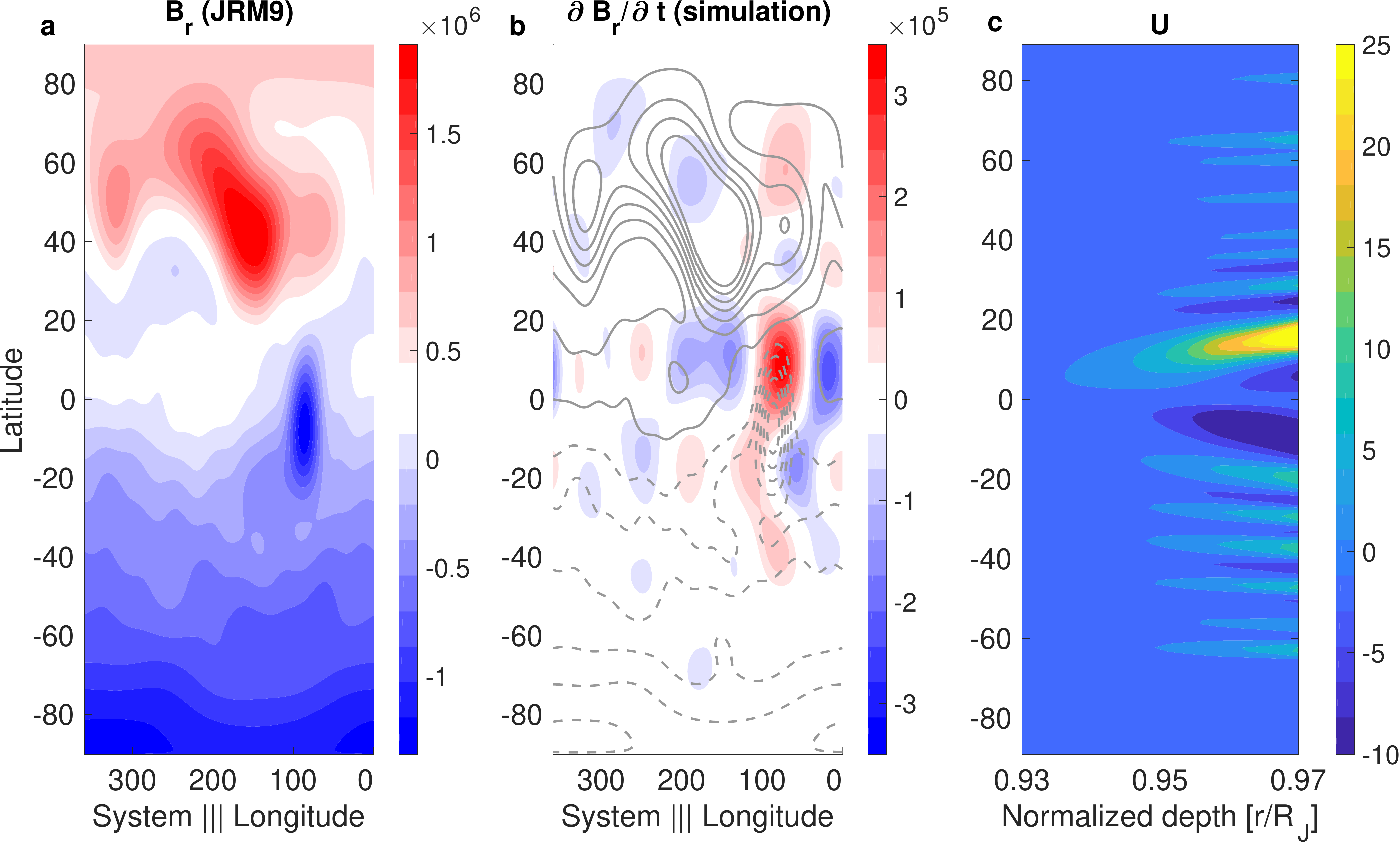}
\par\end{centering}
\caption{\label{fig:Br simulated} (a) The magnetic field $[{\rm nT}]$ measured
from Juno's first nine orbits (JRM9) at $\sim0.972R_{J}$ \citep{Connerney2018}
(b) the changes in the magnetic field $[{\rm nT\cdot yr^{-1}}]$ caused
by the simulated profile after 1 year at the same depth, with gray
contours showing the measured magnetic field (JRM9) and (c) the simulated
zonal flow as function of latitude and depth $[{\rm ms^{-1}}]$ at depths between $0.97\,R_{J}$ and $0.93\,R_{J}$.
Both (a) and (b) are represented here with only 10 degrees and orders
of the magnetic harmonic coefficients as in \citet{Connerney2018}.}
\end{figure}

\begin{table}[h]
\begin{centering}
\includegraphics[width=0.6\textwidth]{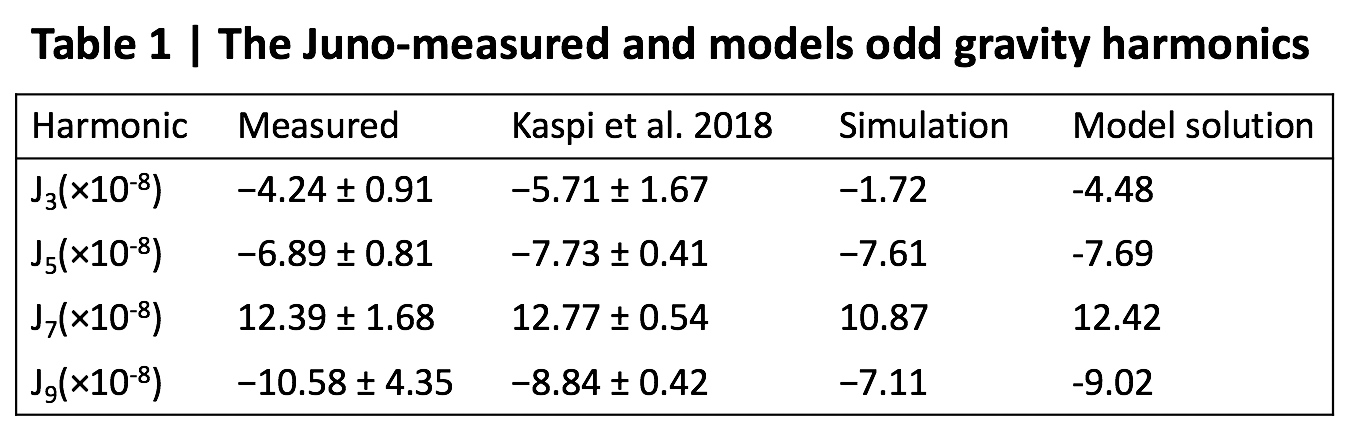}
\par\end{centering}
\label{tab:Jn table} \caption{The measured odd gravity harmonics from Juno, the \citet{Kaspi2018} model
results, the simulated profile and the new combined model results.
The uncertainties are the $3\sigma$ uncertainty values for the measurements
and the \citet{Kaspi2018} model.}
\end{table}
The changes in the radial component of the magnetic field resulting
from the simulated decay profile are shown in Fig. \ref{fig:Br simulated}b,
along with the measured $B_{r}$ from Juno's first nine orbits (Fig.
\ref{fig:Br simulated}a) \citep{Connerney2018}, and the simulated
flow that caused this change (Fig. \ref{fig:Br simulated}c). Both
magnetic fields (Fig. \ref{fig:Br simulated}a,b) are constructed
with only 10 degrees of freedom of the harmonic coefficients l and
m, as in \citet{Connerney2018}. The strongest changes in the magnetic
field appear at latitudes $0-20^{\circ}{\rm N}$ and are caused due
to flows that penetrate areas of high electrical conductivity (Fig.
\ref{fig:Br simulated}c). The cylindrical projection of the surface
winds, parallel to the axis of rotation, causes the strong jet
at $20^{\circ}{\rm N}$ at the surface to shift equatorward and causes
the strong variation in $B_{r}$ at $\sim10^{\circ}{\rm N}$. We will
next consider this time varying magnetic field as an example for the
expected Juno measurements.

\subsection{The combined gravity-magnetic optimization \label{subsec:Definition-of-combine}}

The measurements of the magnetic field, given at the planet's surface,
can be projected in the radial direction to the regions where the
conductivity is no longer negligible \citep{Galanti2017c}, according
to potential field continuation. The comparison between the model
and the measurement should be at this depth, chosen here as ${\rm R}_{c}=0.972R_{J}$
as in \citet{Galanti2017c}. We choose to fit the measurements with
a simple exponential decay flow profile in the regions of the induction
model as we expect that the strong electrical conductivity in this
region will cause the flow to dissipate fast and proportionally to
the electrical conductivity itself which is also exponential. The
flow in this region is modeled such that

\begin{equation}
U_{{\rm Model}}(\theta,r)=U_{{\rm surf}}(s)Q_{M}(r),\label{eq:induction wind}
\end{equation}
\begin{equation}
Q_{M}(r)=U_{M}\exp\left(\frac{r-0.972R_{J}}{H_{M}}\right)\qquad0.93R_{J}\leq r\leq0.972R_{J},\label{eq:induction model decay}
\end{equation}
where $U_{M}$ is the flow strength at $r=0.972R_{J}$ relative to
the surface flow and $H_{M}$ is the exponential decay rate. The search
for the best solution is made by comparing the resulting changes in
the magnetic field from the model to the simulation. Each model run
with different $U_{M}$ and $H_{M}$, using a cost-function (scalar
measure) to find the best solution. The cost-function is defined as

\begin{equation}
L_{{\rm MSV}}=\sum_{n=1}^{10}\sum_{m=0}^{n}\left[\left(\frac{\partial g_{nm}^{{\rm sim}}}{\partial t}-\frac{\partial g_{nm}^{{\rm mod}}}{\partial t}\right)^{2}+\left(\frac{\partial h_{nm}^{{\rm sim}}}{\partial t}-\frac{\partial h_{nm}^{{\rm mod}}}{\partial t}\right)^{2}\right]\qquad0.93R_{J}\leq r\leq0.972R_{J},\label{eq:magnetic cost definition}
\end{equation}
where $g_{nm}^{{\rm sim}}$ and $h_{nm}^{{\rm sim}}$ are the simulated
harmonic Schmidt coefficients \citep{Connerney2018} at the upper boundary of the model and
$g_{nm}^{{\rm mod}}$ and $h_{nm}^{{\rm mod}}$ are the model solution
for those coefficients (not to be confused with the gravity anomaly
$\Delta g_{r}$). We then define the model solution as the average
between all the solutions of the lowest order of magnitude of the
resulting cost-function.

We define the transition between the MSV model and the TW model as
a distinct boundary $\left(R_{T}\right)$, which needs to be set in
the optimization process. For the upper region of the solution $\left(R_{T}<r\leq R_{J}\right)$,
we use the TW inversion approach \citep{Galanti2016,Galanti2017b}, fitting
the gravity odd harmonics that resulted from the forward model to
the measured ones, using a second cost-function,

\begin{equation}
L_{{\rm {\rm TW}}}=\sum_{i=3,5,7,9\,}\sum_{j=3,5,7,9\,}w_{ij}\left(J_{i}{\rm ^{{\rm obs}}}-J_{i}{\rm ^{{\rm mod}}}\right)\left(J_{j}{\rm ^{{\rm obs}}}-J_{j}{\rm ^{{\rm mod}}}\right)\qquad R_{T}\leq r\leq R_{J},\label{eq:gravity cost definition}
\end{equation}
where $J_{n}{\rm ^{{\rm obs}}}$ are the Juno measurements, $J_{n}{\rm ^{{\rm mod}}}$
are the model solutions and $w_{ij}$ is the $4\times4$ weight matrix
as in \citet{Kaspi2018}. In the upper region, we find the best fit
by optimizing the decay profile independently at each vertical grid
point of the model, demanding that the vertical structure function
Q is 1 at $R_{J}$ and has the value of the resulted flow strength
from the MSV model at the transition depth $Q_{M}(R_{T})$. We also
demand that the vertical structure function decreases monotonically
with depth. The decay profile in the region between the transition
depth and the upper boundary of the Induction model $\left(R_{T}\leq r\leq0.972R_{J}\right)$
is recalculated with the TW model. However, this region is still aligned
with considerable conductivity, therefore we must recalculate the
changes in the magnetic field to make sure that the solution is still
compatible with the measurements. The value of the transition depth
between the models $\left(R_{T}\right)$ is determined as deep as
possible with the requirement that the time varying magnetic field
results remain valid.

\section{Results\label{sec:Results}}

\begin{figure}[h]
\begin{centering}
\includegraphics[width=0.8\textwidth]{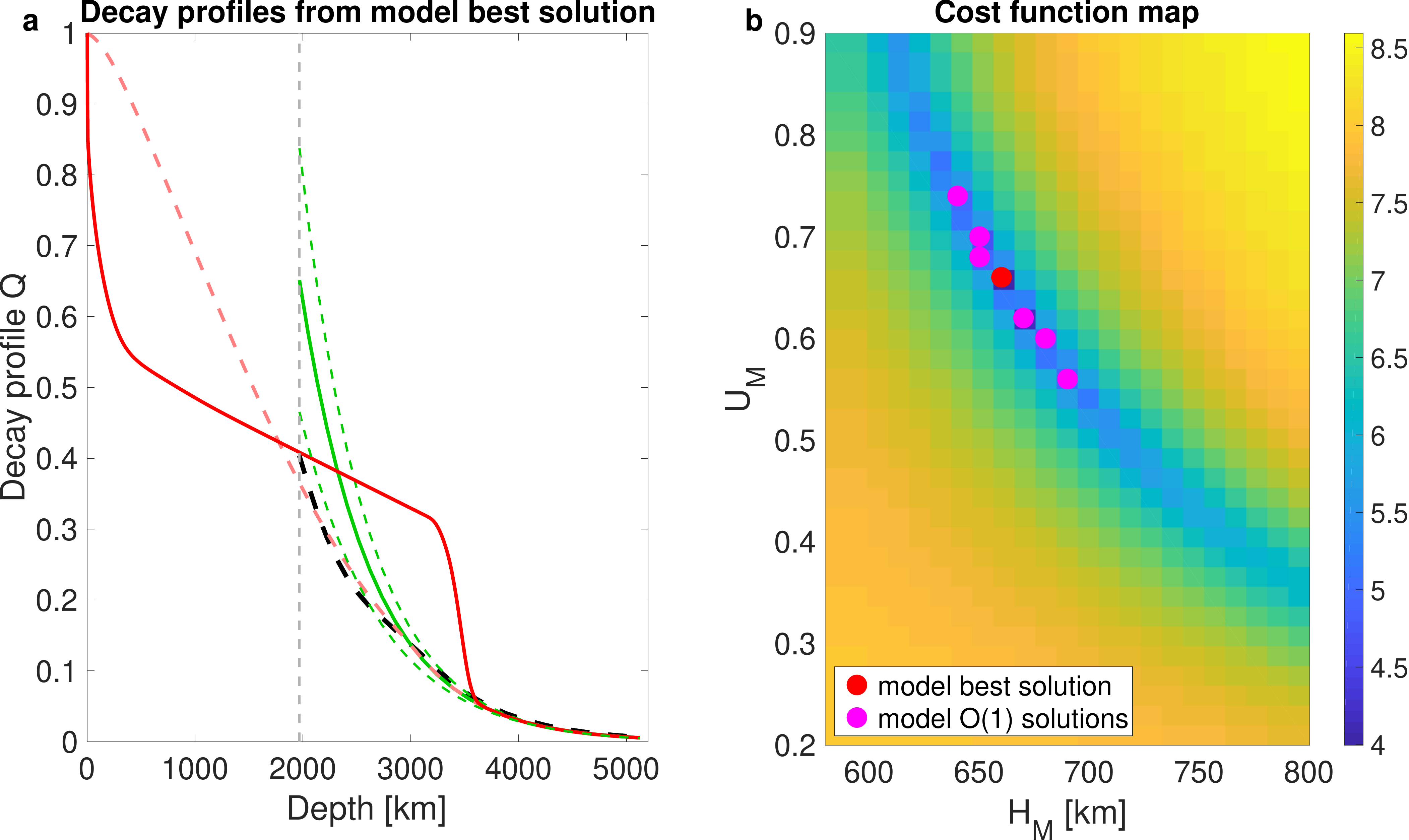}
\par\end{centering}
\caption{\label{fig:model solution and cost} (a) The simulated decay profile
(dashed black) together with the induction model solution for the
lower region (green), its $3\sigma$ uncertainty (dashed green), the
TW model solution for the upper region combining with the Induction
solution below $R_{T}$ (red) and an additional option for the combined
solution (light dashed red); also shown is the induction model boundary
at $0.972\,R_{J}$ (dashed gray). Note that the induction model solution
is very good below $\sim3000$ km, in regions where the conductivity
is high. (b) The cost function map for the induction model as function
of the two optimized parameters: the exponential decay rate $\left(H_{M}\right)$
and the flow strength at the induction model upper boundary relative
to the surface flow $\left(U_{M}\right)$. Also shown are the model
best solution (red and all solutions within O(1) from it (pink).}
\end{figure}

We start with the optimization of the lower region of the decay profile
using the MSV model. The cost function (Eq.~\ref{eq:magnetic cost definition})
shows a near linear relation between the parameters $U_{M}$ and $H_{M}$
(Fig. \ref{fig:model solution and cost}b). Since an optimal solution
in the global minimum is not a unique point but an area (blue region
in Fig. \ref{fig:model solution and cost}b), we average the lowest
order of magnitude of solutions (pink and red dots in Fig. \ref{fig:model solution and cost}b)
to determine the optimal decay rate in the inner layers (Fig. \ref{fig:model solution and cost}a,
green line). The solution is obtained with $U_{M}=0.6514$ and $H_{M}=0.66$.
 The resulting model solution (green line) is very far from the simulated
decay profile (black dashed line) in the region of $2000-3000$ km,
but converges to the simulated decay profile deeper than $\sim3000$ km.
This suggests that the transition depth ($R_{T}$, Eq.~\ref{eq:gravity cost definition})
should be deeper than $\sim3000$ km ($\cong0.958\,R_{J}$),
chosen here as $R_{T}=3500$ km ($\cong0.95\,R_{J}$).

Next, using the TW model, we find a solution for the decay profile
that is constrained with the MSV solution at $R_{T}=0.95R_{J}$. The
new solution fits both the Juno measurements and the simulation for
the changes in the magnetic field (Fig. \ref{fig:model solution and cost}a,
red). This combined gravity-magnetic solution shows a unique pattern
of two rapid and perhaps unrelated decay patterns. The first, close
to the cloud-level, might account for a baroclinic atmospheric outer
layer \citep[e.g.,][]{Kaspi2007} that decays until reaching a nearly barotropic state;
and the second, around $H=3500$ km, that could reflect the magnetic
braking. However, the first decay is very close to the cloud-level
where neither model is sensitive (it is a non-conductive layer and
the density anomalies are small), so this decay is not necessarily
physically meaningful. For example, another more smoother solution can be constructed, that is optimized with less free parameters
(Fig. \ref{fig:model solution and cost}a, light dashed red). This
solution is more similar to the \citet{Kaspi2018} solution, but its
ability to explain all the odd gravity harmonics is not as good as
the best-fit solution (Fig. \ref{fig:model solution and cost}a, red).
The four odd gravitational harmonics associated with this solution
are: $J_{3}=-1.71,\,J_{5}=-6.83,\,J_{7}=9.86$ and $J_{9}=-6.54$.

\begin{figure}[h]
\begin{centering}
\includegraphics[width=0.8\textwidth]{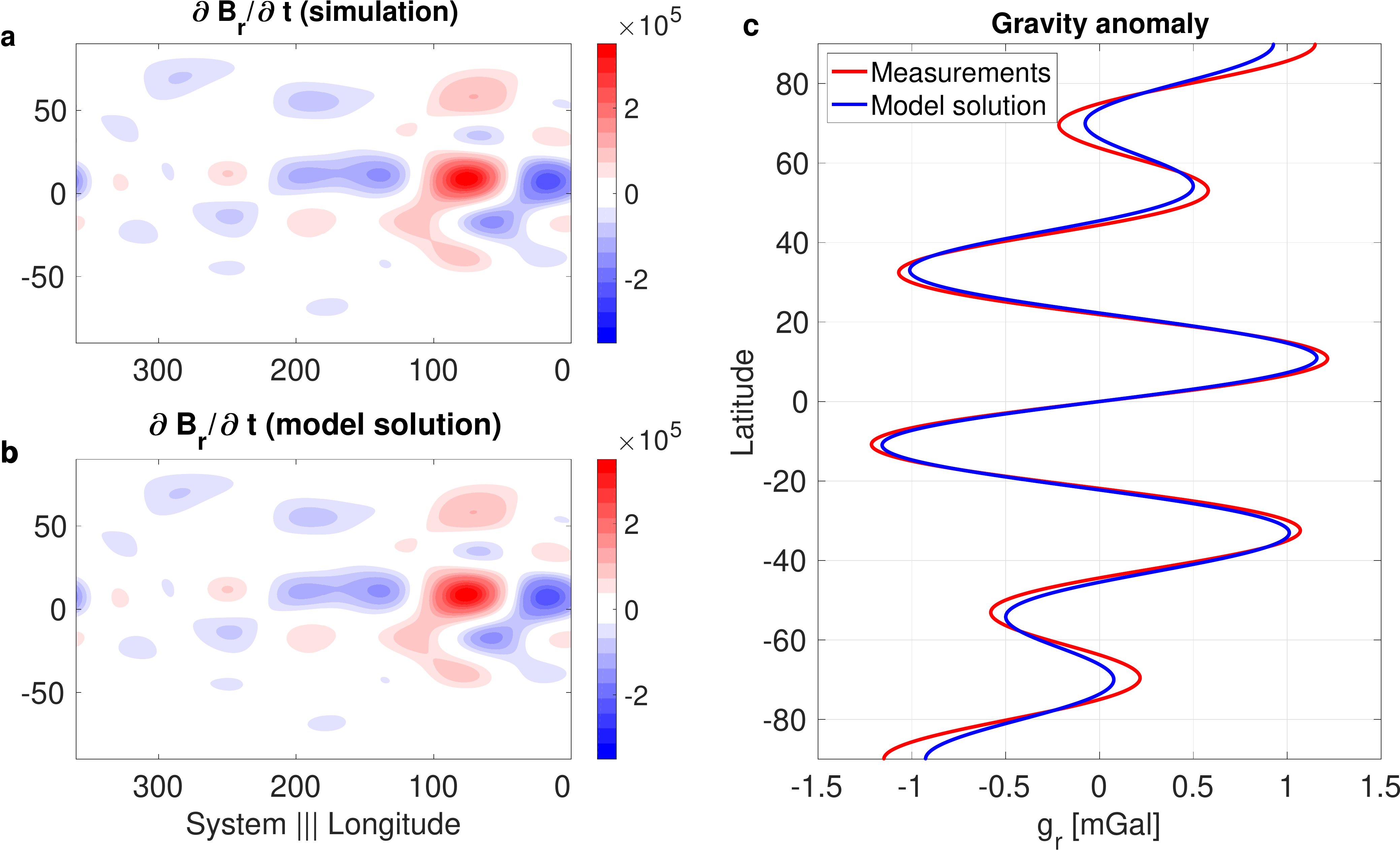}
\par\end{centering}
\caption{\label{fig:measurements compare to model} The combined gravity-magnetic
solution. (a) The resulting time changes in ${\rm B_{r}}$ $[{\rm nT\cdot yr^{-1}}]$
from the simulation, (b) the model solution for the changes in ${\rm B_{r}}$
$[{\rm nT\cdot yr^{-1}}]$ after the same time span, and (c) the gravity
anomalies as measured by Juno (red) and as calculated from the model
best solution result (blue). }
\end{figure}

To make sure the combined best-fit solution fits also the MSV measurements,
the magnetic field anomalies are recalculated from the new profile.
The resulting magnetic field variation is almost identical to the
variation from the MSV model solution alone. This confirms that the
changes in the magnetic field are not sensitive to the flow structure
above $\sim3000$ km ($\cong0.958R_{J}$). The gravity harmonics of
the resulted decay profile are displayed in Table 1 (last column),
the changes in the magnetic field are shown in Fig. \ref{fig:measurements compare to model}b,
and the gravity anomalies in Fig. \ref{fig:measurements compare to model}c
(blue). Fig. \ref{fig:measurements compare to model} also shows the
simulated changes in the magnetic field (a) and the Juno measured
gravity anomalies (c, red). 

The method presented here provides a very good solution for the measurements
created by the simulated profile, but should not be taken as a solution
for Jupiter's deep flows, since the magnetic field changes are based
on our simulation and not the actual measurements. Once the magnetic
measurements based on the following perijovs are
available, the solution for the flow decay profile could be recalculated.
Fitting these two independent measurements simultaneously will help
to better constrain Jupiter's deep flow structure.

\section{Conclusions\label{sec:Conclusions}}

So far, the Juno gravity measurements showed that cloud-level winds
extend to a depth of about $3000$ km \citep{Kaspi2018}. As the gravity
field is mostly sensitive to density anomalies (and hence to flow
strength) at depths of no more than a couple thousand kilometers,
it is not possible to strongly constrain the strength of the flow
below that since the flow there is weak. Measurements of the time
varying magnetic field of Jupiter might help to tackle this problem.

Here, we present a new combined method that uses two independent measurements
from Juno to better resolve Jupiter's internal flow structure. We
use the thermal wind balance that relates the flow structure to the
asymmetric gravity field, and the electromagnetic induction equation
that explains changes in the measured magnetic field caused by the
deep flow. Since the time varying magnetic field measurements are
not available yet, we use the current gravity measurements and simulate
changes in the magnetic field to find a decay profile that fits both.
We find that the flow decay profile of \citet{Kaspi2018} is too strong
at depths of more than $\sim3000\,{\rm km}$ and that further constraints
must be added in order to fit both measured fields. We show that a
decay profile that fits both the gravity measurements and the simulated
magnetic measurements can be found through a joint optimization. Once
the new magnetic measurements will be available we can reoptimize
the solution to find a better flow structure. 

We also find that $J_{3}$ is very sensitive to the deeper regions
of the atmosphere while the MSV method limits the same region to low
velocities, therefore posing upper and lower bounds for the flow
velocity in the deep regions. The MSV method further constrains the
deepest regions of the model such that strong flows cannot exist at
all below $\sim4000\,{\rm km}$, and nearly no flow (less than $1\,{\rm ms^{-1}}$)
can exist below $\sim5000\,{\rm km}$. Finally, we find that the best
solution is characterized with two rapid decays instead of one as
in \citet{Kaspi2018}, which could represent two distinct decay mechanisms
for the flow, each at a different depth range.

\textit{Acknowledgments:}

We thank H. Cao for providing the MSV code and for the very helpful
discussions and suggestions. We also thank J.E.P. Connerney, J. Bloxham
and R. Holme for very useful comments. This research has been supported
by the Israeli Space Agency and the Helen Kimmel Center for Planetary
Science at the Weizmann Institute of Science.

\end{document}